\documentclass[conference]{IEEEtran}
\usepackage{algorithm2e}[linesnumbered]
\usepackage[T1]{fontenc}
\usepackage{booktabs}
\usepackage{cite}
\usepackage{makecell}
\usepackage{multirow}
\usepackage{flushend}
\ifCLASSINFOpdf
  \usepackage{subfig}
  \usepackage[pdftex]{graphicx}
  \graphicspath{{./graphics/}}
  \DeclareGraphicsExtensions{.pdf,.jpeg,.png}
\else
\fi
\begin{document}
\title{Memory-Efficient Dataflow Inference for Deep CNNs on FPGA}

\author{

\IEEEauthorblockN{Lucian Petrica\IEEEauthorrefmark{1}, Tobias Alonso\IEEEauthorrefmark{1}\IEEEauthorrefmark{3}, Mairin Kroes\IEEEauthorrefmark{1}\IEEEauthorrefmark{2}, Nicholas Fraser\IEEEauthorrefmark{1}, Sorin Cotofana\IEEEauthorrefmark{2}, Michaela Blott\IEEEauthorrefmark{1}}
\IEEEauthorblockA{\IEEEauthorrefmark{1}Xilinx Research Ireland, \IEEEauthorrefmark{2}TU Delft, \IEEEauthorrefmark{3}Autonomous University of Madrid}
}

\maketitle

\begin{abstract}

Custom dataflow Convolutional Neural Network (CNN) inference accelerators on FPGA are tailored to a specific CNN topology and store parameters in On-Chip Memory (OCM), resulting in high energy efficiency and low inference latency. However, in these accelerators the shapes of parameter memories are dictated by throughput constraints and do not map well to the underlying OCM, which becomes an implementation bottleneck. In this work, we propose an accelerator design methodology - Frequency Compensated Memory Packing (FCMP) - which improves the OCM utilization efficiency of dataflow accelerators with minimal reduction in throughput and no modifications to the physical structure of FPGA OCM. To validate our methodology, we apply it to several realizations of medium-sized CIFAR-10 inference accelerators and demonstrate up to 30\% reduction in OCM utilization without loss of inference throughput, allowing us to port the accelerators from Xilinx Zynq 7020 to 7012S, reducing application cost. We also implement a custom dataflow FPGA inference accelerator for a quantized ResNet-50 CNN, utilizing on-chip weights, the largest topology ever implemented with this accelerator architecture. We demonstrate that by applying FCMP to the ResNet accelerator, the OCM bottleneck is alleviated which enables the accelerator to be ported from Alveo U250 to the smaller Alveo U280 board with less throughput loss compared to alternative techniques. By providing a finer-grained trade off between throughput and OCM requirements, FCMP increases the flexibility of custom dataflow CNN inference designs on FPGA.
\end{abstract}


%
\IEEEpeerreviewmaketitle

\section{Introduction}

The FPGA implementation of Convolutional Neural Network (CNN) inference accelerators has become a hot topic of research as the efficiency of the inference process increasingly drives the cost of ML-based mobile and datacenter workloads \cite{hazelwood2018applied,wu2019machine}. Modern high-accuracy CNNs for vision applications have deep and complex topologies, consisting of a multitude of convolutional layers \cite{rawat2017deep}\cite{He2015}. While these deep networks have well-established advantages with regard to expressiveness and generalization \cite{zhang2018tropical,Poggio201907369}, they create difficulty for FPGA acceleration due to the large number of parameters which need to be stored and operations to be computed for each inference. 

\begin{figure}[]
\centering
\includegraphics[width=0.45\textwidth]{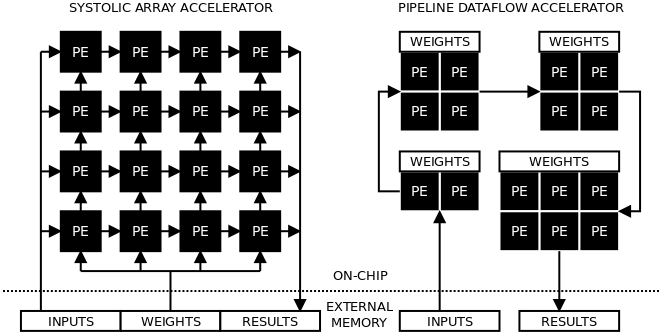}
\caption{FPGA Inference Accelerator Architectures}
\vspace{-0.25cm}
\label{fig:systolic_v_pipeline}
\end{figure}

Figure \ref{fig:systolic_v_pipeline} illustrates two approaches to inference acceleration on FPGA. Most FPGA inference accelerators are based on overlay architectures (Fig. \ref{fig:systolic_v_pipeline} left) \cite{guo2017survey,abdelouahab2018accelerating}, i.e. general-purpose matrix multiplication circuits onto which compute is scheduled to execute the CNN layers in sequence. This approach is flexible, as it enables potentially any CNN topology to be executed by a single accelerator, but is not efficient: it requires frequent transfers of weights and activations (i.e. outputs of hidden layers) between FPGA on-chip scratchpad memories, and external memory (DDR/HBM), arithmetic precision is fixed, and the high execution latency of layer-by-layer compute makes real-time inference difficult.

An alternative FPGA accelerator architecture is custom dataflow (Fig. \ref{fig:systolic_v_pipeline} right), whereby the structure of the FPGA-implemented logic mirrors the topological structure of the CNN, creating a pipeline of smaller compute units, each performing the computation corresponding to one specific CNN layer using weights stored in On-Chip Memory (OCM) and the arithmetic of processing elements is tailored to the precision requirements of each layer. As weights and activations never move off-chip, latency and power dissipation are reduced. Custom dataflow for CNN inference has achieved the lowest latency, highest throughput, and lowest power dissipation for image classification and related tasks up to CIFAR-10 complexity \cite{finn_trets}. However, this accelerator architecture cannot scale to arbitrarily large CNNs, as it is fundamentally limited by available on-chip resources - LUTs and DSP blocks - required to implement compute units for each CNN layers, and critically, the size of OCM required to store the weights. Furthermore, due to mismatches between the sizes of weight buffers and of FPGA OCM blocks, a significant part of the available OCM cannot be utilized for storing weights, further limiting the size of networks which can be accelerated in custom dataflow.

In this work we approach this problem and propose a method to make more of the FPGA OCM available for weight storage in custom pipeline dataflow (onwards denoted simply as dataflow). We present the largest and deepest CNNs implemented to date in single-chip dataflow on FPGA. Using quantization techniques, we develop high-accuracy CNN image classification models based on the ResNet-50 topology, utilizing binary (1-bit) and ternary (2-bit) weights respectively, thus enabling their implementation in existing FPGAs. 

Furthermore, we alleviate the OCM bottleneck of dataflow acceleration as follows. We first modify the popular FINN dataflow accelerator architecture\cite{finn_trets} by utilizing a producer-consumer, Globally Asynchronous Locally Synchronous (GALS) approach for weight storage, whereby memory resources operate at a faster frequency than the compute logic. By leveraging the memory to compute frequency ratio $R_F$, we can multiplex the two available RAM ports on Block RAMs (BRAMs) in Xilinx FPGAs to expose $2R_F$ virtual RAM ports to the compute logic. We subsequently utilize a bin packing algorithm to pack parameter memories in groups of size up to $2R_F$, such that each group optimally fills one BRAM, and each member of the group can be read in every compute cycle. This design methodology, which we call Frequency Compensated Memory Packing (FCMP), improves the OCM utilization with minimal reduction in inference throughput and no modifications to the physical structure of the FPGA fabric.

To validate our methodology, we apply it to BNN-Pynq\footnote{https://github.com/Xilinx/BNN-PYNQ}, a collection of FINN-generated medium-sized CIFAR-10 inference accelerators for binarized neural networks, targeting Zynq FPGA devices, as well as ResNet-50 targeting Alveo accelerator cards. Our specific contributions are:

\begin{itemize}
    \item We describe two quantized ResNet-50 CNN models, trained for binary and ternary weights and optimized for FPGA dataflow execution, which achieve 87.64\% and 89.38\% ImageNet Top-5 accuracy respectively, and the highest compute density to date for models amenable to dataflow execution on a single FPGA. We demonstrate over 2700 FPS and under 2ms latency for the binary-weight ResNet-50 on Alveo U250.
    \item We describe modifications to the FINN dataflow architecture and an accelerator design methodology, which utilizes FCMP to maximize the efficiency of OCM utilization for CNN dataflow accelerators at small or large scales.
    \item We apply the previously described methodology to the BNN-Pynq and ResNet accelerators and achieve up to 30\% OCM reduction with a 5\% LUT utilization overhead, and at most 32\% decrease in top frequency. This reduction in OCM utilization opens up opportunities for implementing the accelerators in smaller devices than previously possible. The technique enables us to implement the binary ResNet50 in Alveo U280 at the same folding factors used on Alveo U250, and shifts the design bottleneck of the ternary ResNet-50 from OCM to LUTs.
\end{itemize}

We provide background on FPGA dataflow inference and OCM optimization techniques in the following section and describe our dataflow accelerators for ResNets in Section \ref{sec:resnet}. We describe our frequency compensated memory packing methodology in Section \ref{sec:Packing}, and evaluate it in Section \ref{sec:Evaluation}.

\section{Background and Previous Work}

\subsection{FPGA Dataflow Inference of Quantized CNNs}

Recent work on techniques for neural network binarization \cite{courbariaux2016binarized, zhou2016dorefa} has reduced the computational and memory requirements of NN inference, enabling the implementation of multiple dataflow accelerators and accelerator frameworks \cite{finn_fpga17, finn_trets, ghasemzadeh2018rebnet} for binarized and quantized NNs (QNNs) in FPGA. However, most dataflow accelerators described in previous work target relatively small binarized CNNs which achieve acceptable accuracy on simple image and text processing tasks, e.g. classification for MNIST, CIFAR-10, SVHN datasets in \cite{finn_fpga17} and character recognition in \cite{rybalkin2018finn}. Dataflow-style FPGA-accelerated binarized CNNs for the ImageNet \cite{deng2009imagenet} 1000-class classification problem have been developed utilizing the FINN \cite{finn_trets} and ReBNet \cite{ghasemzadeh2018rebnet} accelerator frameworks, but have limited Top-1 accuracy in the range of 40-50\%. Recent work in \cite{customjpegresidual} increases Top-1 accuracy of dataflow accelerators to 70\%, but is still lower compared to equivalent GPU and CPU inference solutions and even overlay-style FPGA inference. 

To date, achieving state of the art accuracy with dataflow accelerators in FPGA remains a challenge. While approaches such as utilizing higher weight precision, e.g. 2-bit ternary quantization \cite{li2016ternary}, or deeper NNs such as ResNet-50 \cite{He2015} have the potential to increase achievable accuracy, they also significantly increase the size of the required weight storage, making dataflow acceleration difficult.

\subsection{Memory Efficiency in Dataflow FPGA Inference of CNNs}

Modern FPGA devices advertise large amounts of OCM but most of it is implemented as RAM blocks embedded within the fabric, of a fixed shape and number of ports with typically narrow and deep aspect ratio, e.g. 18 Kb (18b wide and 1024-deep) 2-port memories in Xilinx FPGAs. Efficiently mapping the arbitrarily shaped weight memories of a CNN dataflow accelerator to block RAM resources on FPGA is challenging, due to mis-matches between desired and available memory shapes. Often, most of the Block RAMs (BRAM) are not utilized to their full capacity, and therefore, the actual usable OCM is much less than the device maximum. As a result, OCM is often the bottleneck in dataflow acceleration, as illustrated in Table \ref{tab:bnnpynq_bottleneck} for the accelerators in the BNN-Pynq collection. Note how the BRAM utilization is highest of all resource types.

\begin{table}[]
\small
\caption{\label{tab:bnnpynq_bottleneck}Resource Utilization of FINN Dataflow Accelerators (BNN-Pynq) on Zynq 7020}
\centering
\begin{tabular}{ccccc}
\toprule
\textbf{Accelerator} & \textbf{\makecell{BRAM\\(\%)}} & \textbf{\makecell{LUT\\(\%)}} & \textbf{\makecell{DSP\\(\%)}} \\\midrule
CNV-W1A1        & 88    & 49       &  11  \\
CNV-W1A2        & 94    & 76       &  12  \\
CNV-W2A2        & 100   & 70       &  15  \\
LFC-W1A1        & 78    & 53       &  2  \\
LFC-W1A2        & 79    & 92       &  2  \\
\end{tabular}
\end{table} 

We can distinguish three causes for the inefficient use of FPGA OCM: accelerator folding, convolution kernel sizes, and pruning.

\begin{figure}[]
\centering
\includegraphics[width=0.3\textwidth, trim={9cm, 4.5cm, 10cm, 4cm}, clip=true]{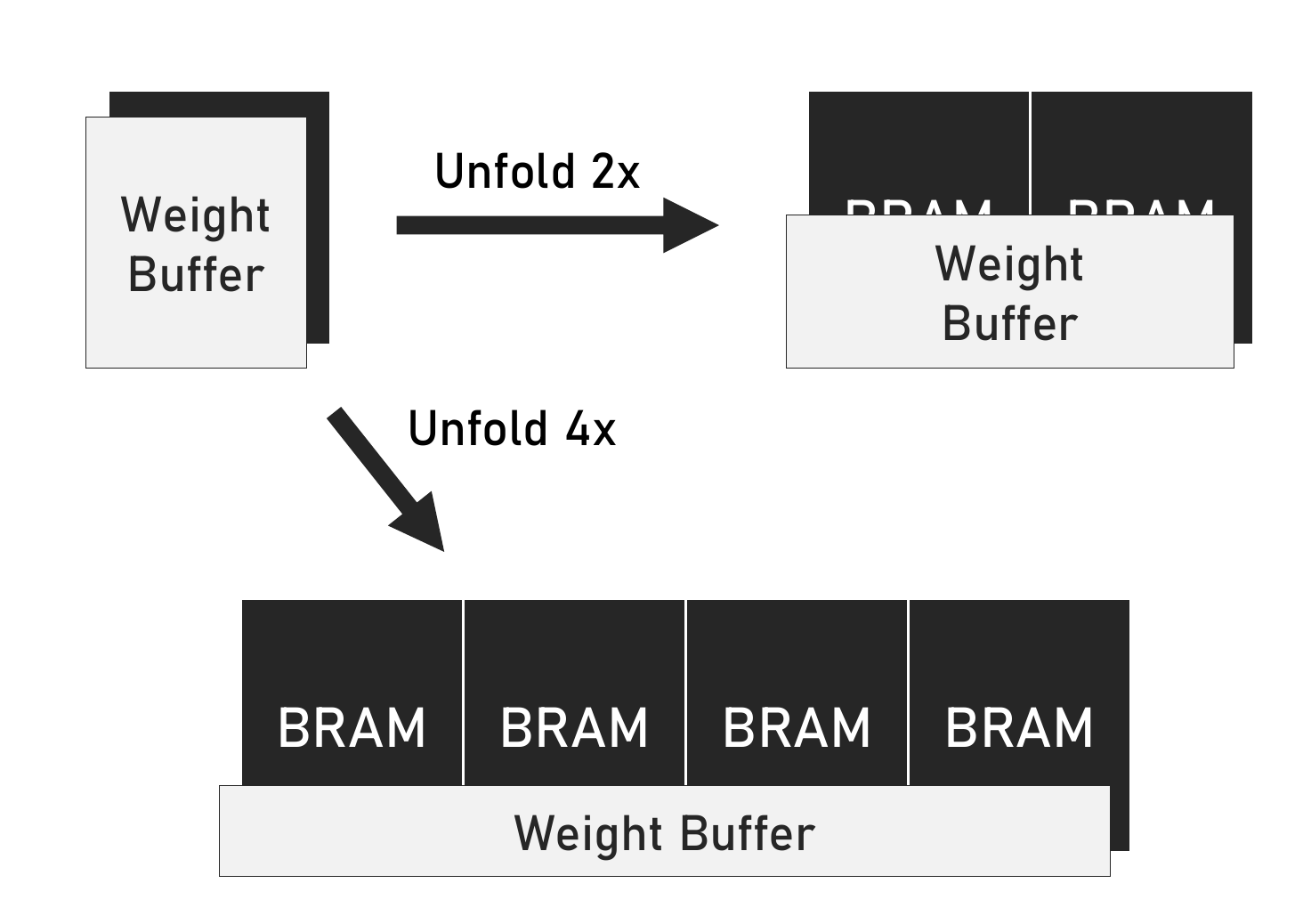}
\caption{Efficiency Decreases with Increased Parallelism}
\vspace{-0.25cm}
\label{fig:efficiency_v_simd}
\end{figure}

\paragraph{Throughput Requirements}
For FINN-style accelerators in particular, the computational throughput can be scaled by folding, i.e allocating a variable number of vector Processing Elements (PEs) to each layer, and scaling the SIMD vector length of the PE. Previous work  has demonstrated this approach enables FINN accelerators to scale to various sizes of FPGA. However, the folding factor not only affects compute logic but also determines shapes (widths, depths) on the parameter memory, in order to achieve parameter readback at the same rate as the compute. Figure \ref{fig:efficiency_v_simd} illustrates this effect - as we double or quadruple the compute capability, we utilize more BRAMs and fewer words of each BRAM. Therefore, despite the total number of parameters being constant, the number of block RAMs scales proportionally to the compute resources of the accelerator.

\begin{equation}\label{eq:efficiency}
E=\frac{N_{p}\cdot W}{N_{RAM}\cdot C_{RAM}}
\end{equation}

We define the physical RAM mapping efficiency as in Equation \ref{eq:efficiency}, where W is the bitwidth of the NN weights, and $N_{p}$ is the number of parameters. $C_{RAM}$ is the total capacity of the RAM blocks required to implement the weight buffer, which depends on the memory width given by the product of $N_{PE}$ and $N_{SIMD}$, the parallelism parameters of the accelerator, and the depth of the parameter memory (see \cite{kroes2020pack} for an exact analytic expression). As parallelism increases, memory depth is reduced, width is increased, increasing $N_{RAM}$ and decreasing efficiency. 

\paragraph{Convolution Kernel Sizes}
Odd-sized convolutional kernels are ubiquitous in deep learning, especially 3x3 kernels (K=3), but also 5x5 (K=5) and larger.
In FINN accelerators, weight buffer depth is proportional to the square of the kernel dimension. We therefore arrive at buffer depths which are multiples of $K^2$, with the resulting efficiency being at most $K^2/2^{ceil(log2(K^2))}$. This measure is lowest for the very popular 3x3 kernel utilized in all modern CNN topologies, and highest for the 1x1 (pointwise) convolution.

\paragraph{Filter Pruning}
Filter pruning techniques \cite{pruning2019gate} eliminate redundant filters from a CNN thereby reducing the total compute required for inference as well as the total number of weights. For FPGA dataflow inference, this reduction in the number of filters for a given layer reduces the depth of the weight buffer. In most cases, this will only lead to an increase in OCM utilization inefficiency, instead of an actual reduction in BRAM utilization. Therefore, the benefits of pruning cannot be fully utilized.

\subsection{Buffer to BRAM Packing}
The problem of efficient mapping of logical buffers to block RAMs has been approached in MPack \cite{chow_mempack} and MemPacker \cite{karchmer1994pack}. MemPacker is based on a branch-and-bound algorithm, while MPack utilizes a simulated annealing approach to discover a good mapping of multiple logical buffers in a single block RAM. MPack supports both vertical co-location (a word from each buffer are concatenated into a physical RAM word) or horizontal (the first buffer occupies the lower addresses, then subsequent buffers start at the next address after the previous buffer ends). MPack is demonstrated on relatively small examples compared to modern inference accelerators, while MemPacker has a high worst-case time complexity. An alternative packing methodology is described in \cite{kroes2020pack} based on genetic algorithms, which is able to perform the packing in a matter of seconds for FINN-style FPGA dataflow inference accelerators up to and beyond the size of the ResNet-50 under analysis in our work. 

While the results in \cite{kroes2020pack} indicate an over 30\% increase in OCM mapping efficiency is possible for ResNet-like topologies using the FINN dataflow architecture, their proposed methodology, as well as that in \cite{chow_mempack}, suffer from an associated reduction in the perceived readback throughput, proportional to the maximum number of logical buffers sharing a BRAM port. For example, each of 4 buffers packed into a dual-port RAM is read back once every 2 cycles, compared to packing 2 buffers into the same RAM, which allows readback of each buffer in every cycle. A potential solution is described in \cite{ullah2018efficient} wherein a Block RAM is overclocked relative to the surrounding computation logic, allowing its ports to be shared without loss of throughput. Our approach is a combination of these techniques and in the remaining sections we evaluate the feasibility of its application to dataflow CNN acceleration as well as the costs associated with it in terms of resource utilization and achievable frequencies.

\section{ResNet-50 Dataflow Inference on FPGA}\label{sec:resnet}

ResNet-50 \cite{He2015} is a deep CNN which achieves high classification accuracy on the ImageNet \cite{deng2009imagenet} benchmark. Unlike previously dataflow-acclerated CNNs, ResNet-50 has a non-linear topology consisting of branch-and-join structures called Residual Blocks (ResBlocks) of 2 types, consisting of 3 or 4 convolutions respectively, with 3 convolutions on one branch and one convolution optionally present on the second branch. In each ResBlock, one convolution utilizes a 3x3 kernel while the others are 1x1 kernels. The entire ResNet-50 consists of 16 such daisy-chained ResBlocks and additional layers at the top (input) and bottom (output) of the network.
In 4 of the 16 ResBlocks, a doubling of the feature map channels also occurs, resulting in an increase from 64 to 1024 feature map channels in the ResBlock sequence.
Consequently, this results in an increase in parameter memory size and overall memory requirement for ResBlocks in the bottom part of the NN compared to the ones at the top.

\subsection{ResNet-50 Quantization}

To enable the FINN-compatible quantization of the network, we reimplemented the original ResNet-50 v1.5 topology in PyTorch \cite{ketkar2017introduction} utilizing quantized convolutions and replacing ReLU activation layers with quantized activation layers from the Brevitas\footnote{https://github.com/Xilinx/brevitas} quantization-aware training library. The weights of the first and last layers were quantized to signed 8 bit integers. We developed two quantized ResNet50 versions, with weights of convolutions within ResBlocks quantized to either 1 bit (binary) or 2 bits (ternary) respectively. In both models, activations leading into and out of the elementwise addition are quantized to 4 bits signed integers, and all other activations are quantized to 2 bits signed integer. Floating-point batch normalization layers are added before every quantized activation layer.
The floating-point scale factors utilized by each activation layer are learned during the quantization-aware training process, using the technique proposed by Esser~et~al.~\cite{esser2019learned} and Jain~et~al.~\cite{jain2019trained}, while making sure to maintain the same scaling factor on all inputs to each elementwise addition.

This particular configuration of the topology achieves a Top-1/Top-5 accuracy of 67.27/87.64 percent for binary weights and 69.85/89.38 percent for ternary weights, and allows the resulting quantized model to be streamlined using FINN. 

\begin{figure}[]
\centering
\includegraphics[trim={5cm, 0.5cm, 5cm, 1cm}, clip=true, width=0.45\textwidth]{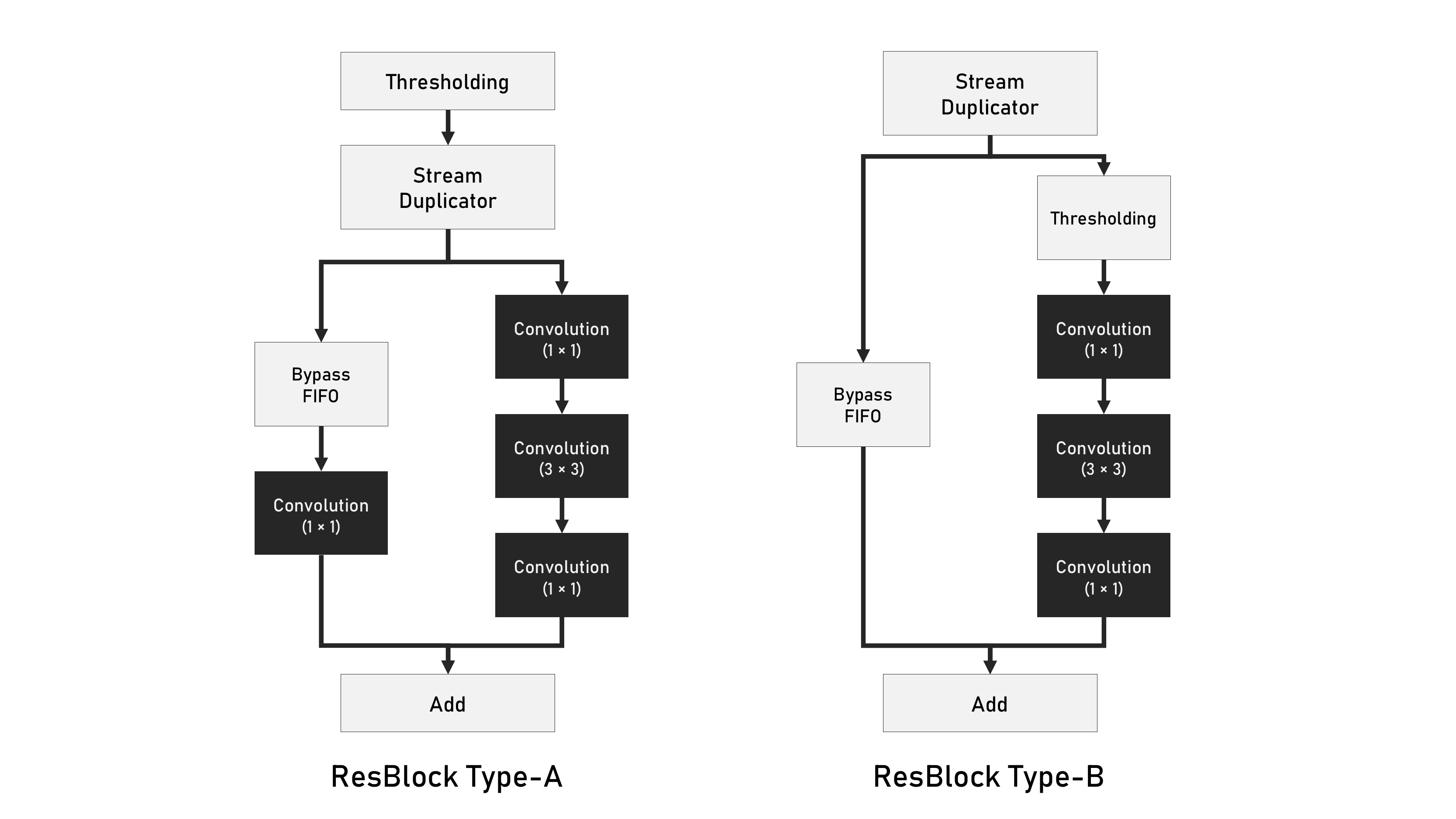}
\caption{Residual Block Structure}
\vspace{-0.25cm}
\label{fig:resblock_structure}
\end{figure}

\subsection{ResNet-50 Dataflow Implementation}

In the FINN streamlining process, batch normalization and quantized activation functions are merged into thresholding operations, which are subsequently merged with quantized convolutions into FINN Matrix-Vector-Activation (MVAU) components which can be implemented in FPGA. The streamlined ResBlocks are illustrated in Figure \ref{fig:resblock_structure}. Utilizing the MVAU resource usage modelling described in \cite{finn_trets}, we developed a folding solution (i.e. values for PE and SIMD parameters for each MVAU) for the binary ResNet-50 which maximizes throughput within the resource limitations of the Alveo U250 FPGA, the largest currently available Xilinx FPGA accelerator card. This modelling exercise indicates that the OCM utilization efficiency for both ResNet-50 models is slightly above 50\%, and OCM is the implementation bottleneck, as expected from dataflow CNN accelerators in the literature.

We utilize  the FINN HLS library\footnote{https://github.com/Xilinx/finn-hlslib} to implement the resblock convolutions as MVAU dataflow blocks, illustrated in Figure \ref{fig:resblock_structure} in black. Other components were implemented in custom C++ code for stand-alone thresholding, stream duplication and elementwise addition, then added to the FINN HLS library. A relatively deep FIFO is required on the bypass path of the resblock to store activations temporarily. The C++ implementation allows setting the precision of stored weights, as well as the values of the PE and SIMD parallelism parameters of the FINN MVAU. This enables the same code-base to support both the binary and ternary variants of the ResNet-50, as well as multiple folding solutions. 

\begin{figure}[]
\centering
\includegraphics[width=0.475\textwidth]{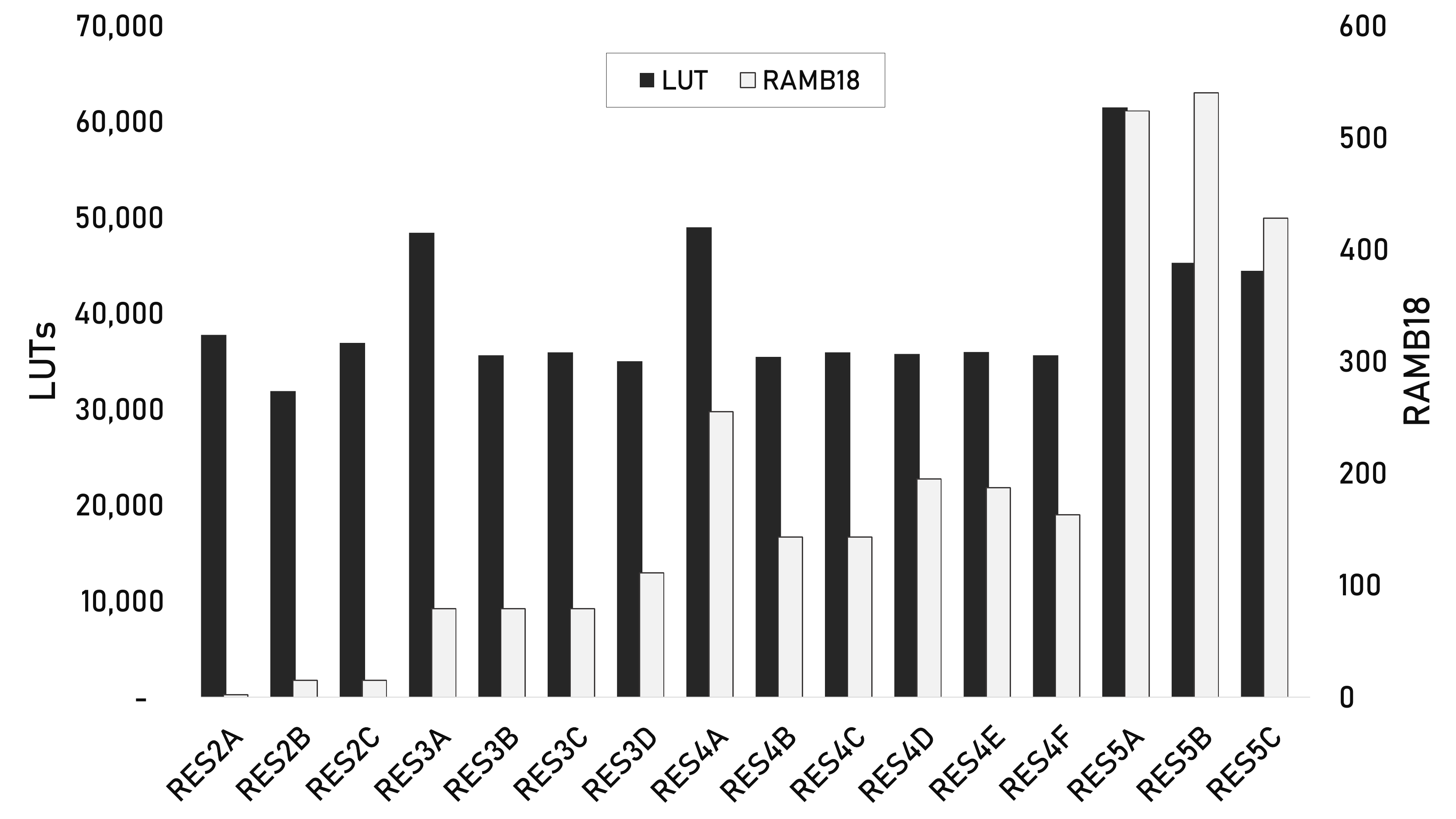}
\caption{ResNet-50 Resource Utilization per ResBlock}
\vspace{-0.25cm}
\label{fig:resblock_utilization}
\end{figure}

\begin{figure}[]
\centering
\includegraphics[trim={1cm 2.5cm 1cm 2.5cm}, clip=true, width=0.475\textwidth]{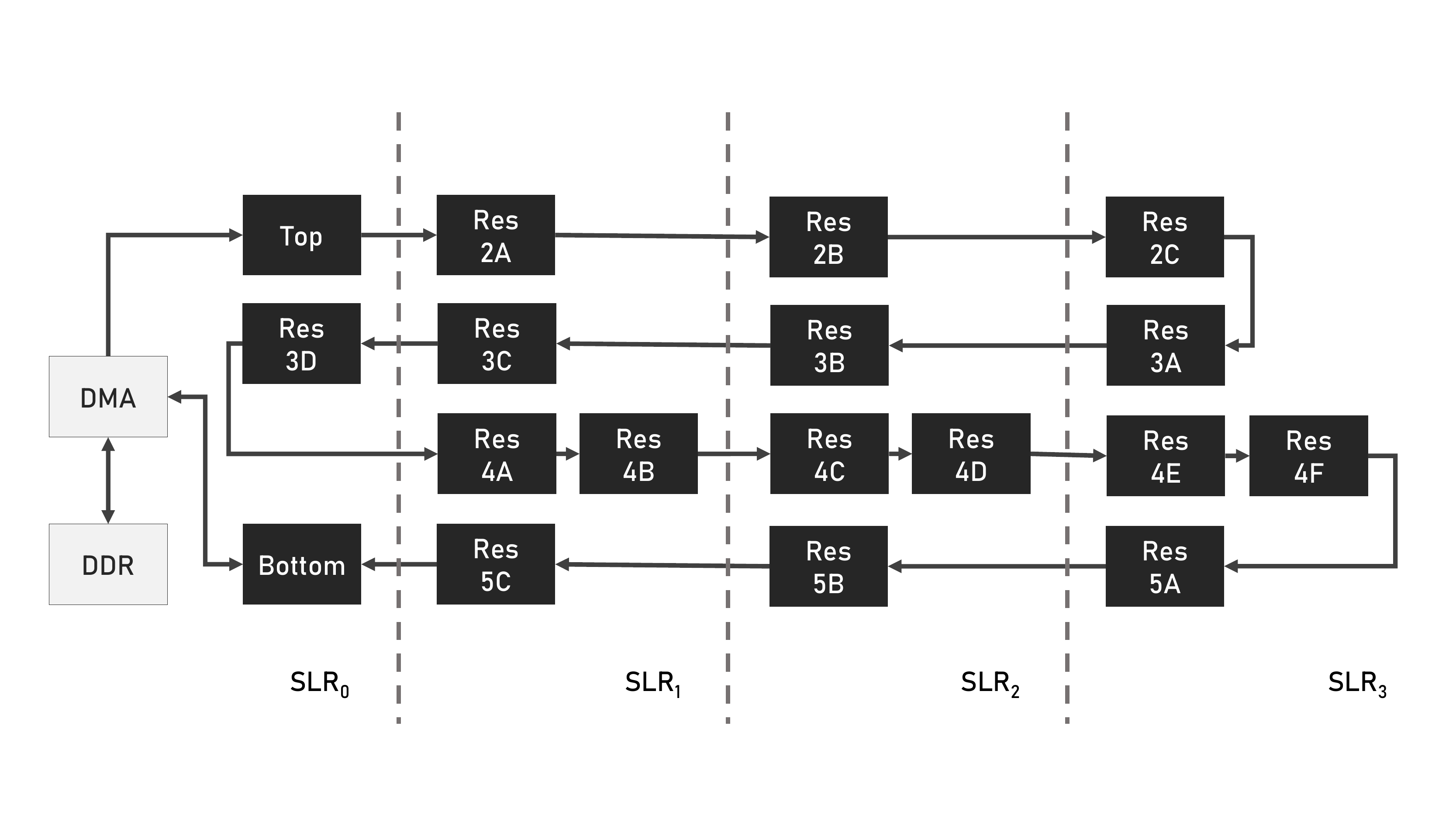}
\caption{ResNet-50 Floorplan on Alveo U250}

\label{fig:resnet_floorplan_u250}
\end{figure}

\begin{table*}[]
\caption{Comparison of Selected FPGA Dataflow Accelerators for ImageNet\label{tab:acc_comparison}}
\centering
\begin{tabular}{ccccccccccc}
\toprule
\textbf{Accelerator} & \textbf{\makecell{Acc. \\ (Top-1 \%)}} & \textbf{TOp/s} & \textbf{\makecell{FPGA \\ Platform}} & \textbf{\makecell{$F_{max}$ \\ (MHz)}} & \textbf{kLUTs} & \textbf{BRAM18s} & \textbf{DSPs} & \textbf{\makecell{Max \\FPS}} & \textbf{\makecell{Min \\ Latency (ms)}} & \textbf{\makecell{Max \\ Power (W)}} \\ \midrule
DoReFaNet-DF\cite{finn_trets}               & 50                  & 11.4           & AWS F1                 & 155           & 477     & 1332      & N/A      & 5241 & N/A & N/A   \\
ReBNet Arch3\cite{ghasemzadeh2018rebnet}    & 41                  & N/A           & VCU108          & 200           & 188     & 3125      & 768      & 170-520 & N/A & N/A   \\
ShuffleNetV2-W1A8\cite{customjpegresidual}    & 70.8                  & 2.42           & AWS F1          & 300           & 274     & 2746      & 2370      & 3321 & N/A & 75   \\
RN50-W1A2 (ours)                       & 67.3                  & 18.3           & Alveo U250             & 195           & 1027     & 3870      & 1611     & 2703  & 1.9 & 71  \\\bottomrule
\end{tabular}
\end{table*}

We synthesized the residual blocks individually with Vitis 2019.2 to evaluate resource usage when targeting Xilinx UltraScale+ FPGAs and to validate the folding solution. Figure \ref{fig:resblock_utilization} indicates that LUT utilization is approximately constant for all ResBlocks but memory utilization increases dramatically towards the output of the network, proportionally with the number of channels. The unbalanced memory utilization creates placement and routing difficulties - the floorplan for the ResNet-50 on Alveo U250 in Figure \ref{fig:resnet_floorplan_u250} is required to fit the accelerator into the device at the chosen folding solution. In our implementation we utilize mostly BlockRAM for weight storage and UltraRAM for activation storage, FIFOs and weights of the final fully connected layer in the ResNet50 topology.  

Table \ref{tab:acc_comparison} presents a comparison between the binary ResNet-50 accelerator implemented on Alveo U250 and dataflow FPGA accelerators in previous work targeting ImageNet classification. Throughout the remainder of this paper we utilize the RN50 prefix to denote ResNet-50 accelerators and a suffix of form WxAy to denote an accelerator which utilizes x-bit weights and y-bit activations. Compared to previous FINN-based work, RN50-W1A2 has a 17\% higher Top-1 accuracy on ImageNet. The resource utilization is higher for our accelerator and the FPS lower compared to DoReFaNet-DF, a fact  explained partially by the difference in total work required (18 TOp/s for RN50-W1A2 vs 11 TOp/s for DoReFaNet-DF), and partially by the additional complexity of the Resnet topology. Compared to \cite{customjpegresidual}, our ResNet designs have slightly lower inference accuracy and throughput, slightly lower power, and a 25\% higher FPS/MHz, perhaps due to being implemented in a larger FPGA card. A variation of the RN50-W1A2 design is available online\footnote{https://github.com/Xilinx/ResNet50-PYNQ}.

LUT-wise, the binary ResNet-50 is small enough to implement in the Alveo U280, the next smallest card in the Xilinx range, but at the default OCM efficiency it requires too many RAMs. Using additional folding, the same network is able to fit the U280 at a 2x throughput penalty. Similarly, the ternary ResNet-50 LUT count is within the capacity of the U250, but its memory requirements are too great, therefore it also requires 2x folding and equivalent reduction in throughput according to the FINN model. This raises the possibility that an increase in OCM utilization efficiency would enable the implementation of the binary ResNet-50 on the smaller Alveo U280 and the ternary on U250 without requiring additional folding of the networks and the associated loss of throughput. In effect, for these two designs and their target FPGAs, an increase in OCM efficiency results in up to 2x increase in throughput. The next section proposes a methodology for increasing this efficiency. 

\section{Frequency Compensated Memory Packing}\label{sec:Packing}

Our guiding insight is that in low-precision FPGA dataflow accelerators, unlike overlay accelerators, the computational logic is typically clocked at relatively low frequencies, in the range of 100 to 300 MHz. This is partly because the complex computation logic is implemented in LUTs rather than DSP tiles, and also because unlike overlays, dataflow accelerators are not usually hand-optimized, but rather are compiled directly from C++ by FPGA design tools. Conversely, memory resources utilize Block or Ultra RAM fabric primitives which are specified for operation at much higher speeds, over 600 MHz. We therefore conjecture that in most dataflow accelerator designs, memory resources are capable of significantly higher maximum frequency than compute logic. Since this higher memory speed cannot increase accelerator throughput, which is limited by the compute speed, we aim to utilize the speed differential to increase efficiency in storing CNN parameters to FPGA OCM. 

The proposed clustering methodology consists of two complementary aspects. First, we employ asynchronous logic design techniques to maximize the read throughput of FPGA OCM, from the perspective of the compute logic. Secondly, we utilize a bin packing algorithm to discover optimal allocations of parameter memories to physical BRAMs, maximizing mapping efficiency, such as the one in \cite{kroes2020pack}.

\begin{figure}[]
\centering
\includegraphics[trim={1cm, 0.5cm, 1cm, 2cm}, clip=true, width=0.5\textwidth]{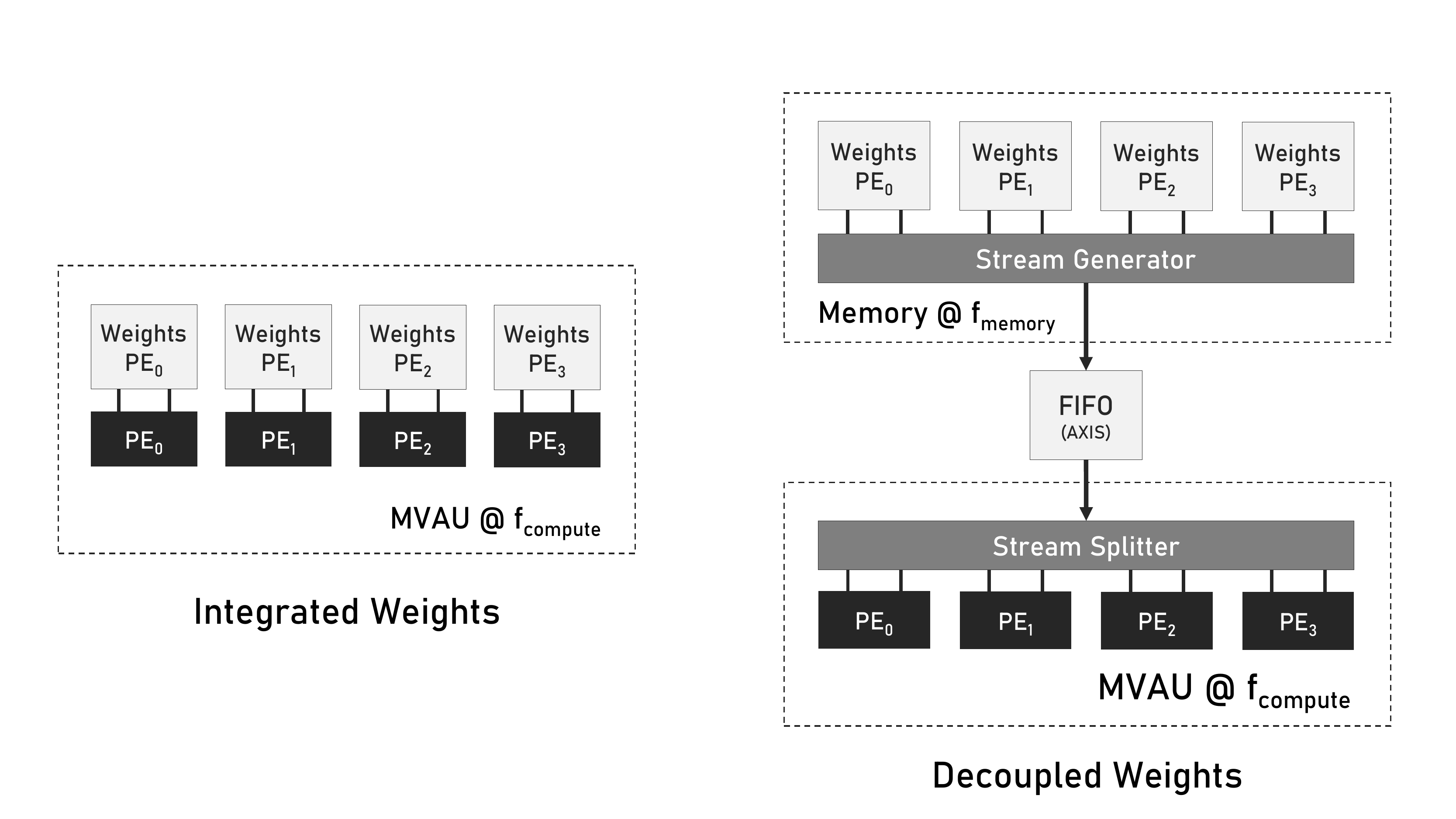}
\caption{GALS Design with Overclocked Memory}
\label{fig:partitioned_design}
\end{figure}

To enable packing a number of partitions greater than the number of physical ports in each BRAM, we make two modifications to the FINN dataflow architecture. First, we separate each MVAU into two blocks: a weight storage block, which also reads the weights in a specific schedule to generate a stream of weight values, and a computational block, which performs the matrix-vector multiplication and activation. We further partition the design into two globally asynchronous but locally synchronous (GALS) islands, corresponding to the weight storage and compute respectively. Weights are transported from the memory storage clock domain to the compute clock domain through AXI Stream interfaces and corresponding asynchronous FIFOs. The original MVAU design as well as the GALS MVAU are illustrated in Figure \ref{fig:partitioned_design}. 

Given the lower complexity of the memory read logic compared to the computation logic, we assume we can clock the memory domain at higher frequency than the compute domain and multiplex the physical memory ports at run-time to read out more than 2 buffers from each physical RAM. The relationship between the number of buffers we can allocate to each physical BRAM without affecting throughput is given by Equation \ref{eq:bin_height}, and we denote the frequency ratio as $R_F$.

\begin{equation}\label{eq:bin_height}
H_B\leq N_{ports}\cdot\frac{F_{memory}}{F_{compute}}
\end{equation}

\begin{figure}[]
\centering
\subfloat[][Example Packing for $R_F=2$]{
\includegraphics[trim={10cm, 2cm, 10cm, 2cm}, clip=true, width=0.23\textwidth]{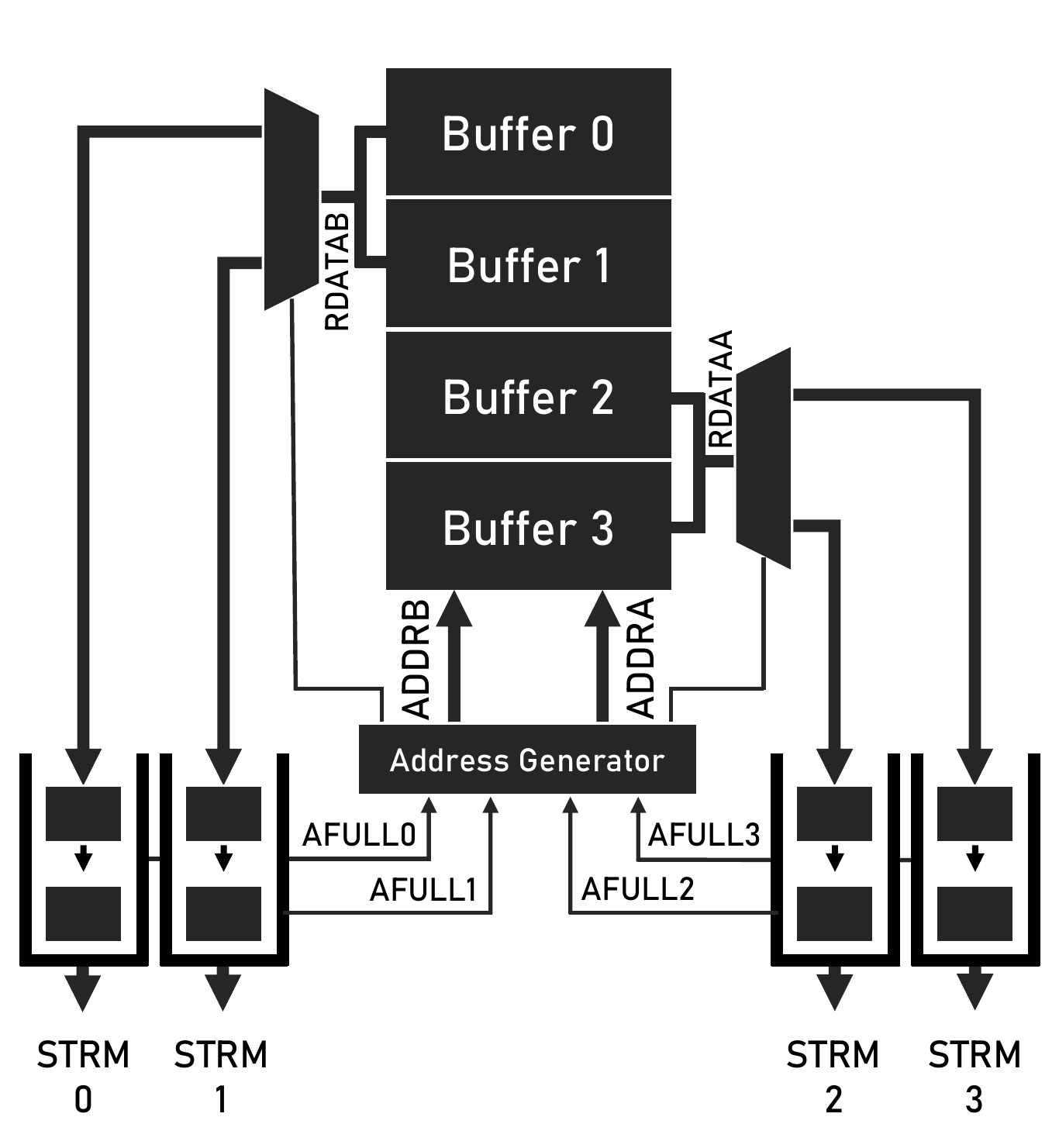}\label{fig:integer_stream}}
\hfill
\subfloat[][Example Packing for $R_F=1.5$]{
\includegraphics[trim={10cm, 0.75cm, 10cm, 2cm}, clip=true, width=0.23\textwidth]{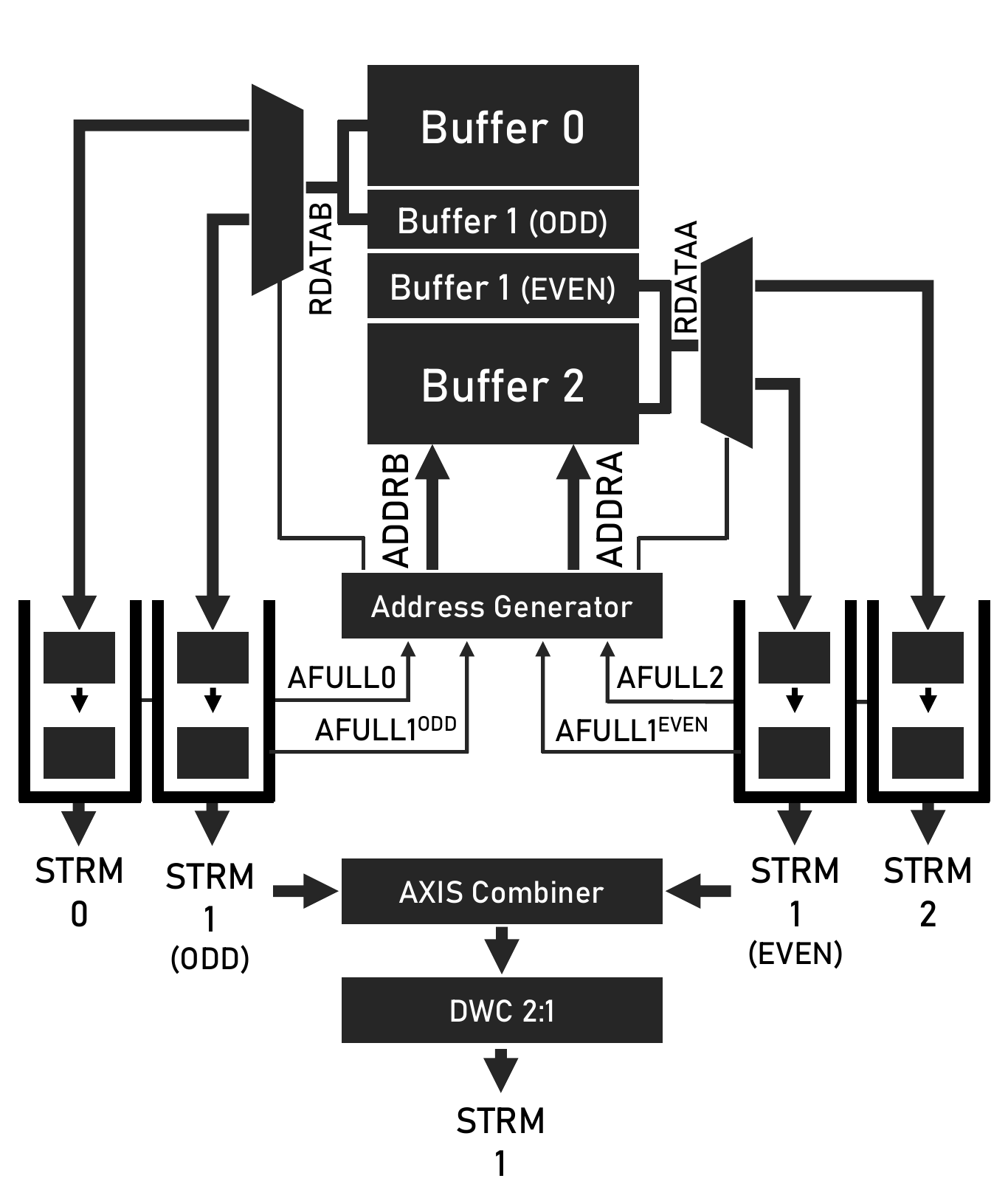}\label{fig:fractional_stream}}
\caption{Streamers with round-robin Port Multiplexing}
\label{fig:streamer_sym}
\vspace{-0.25cm}
\end{figure}

Figure \ref{fig:streamer_sym} illustrates two examples of multiple buffers co-located in a physical BRAM and accessed through port multiplexing in round-robin fashion. In the simplest case we can co-locate an even number $N_b$ of buffers in a BRAM, half of which are served by port A and the other half by port B. This use-case allows for integer frequency ratios $R_F$. At run-time, each of the co-located buffers is read $2/N_b$ times per clock cycle. From the compute logic's perspective, each of the buffers is read $2R_F/N_b$ times per clock cycle. To maintain the compute throughput, we require $R_F \geq N_b/2$. This case is exemplified in Figure \ref{fig:integer_stream}, for a scenario with 4 buffers packed into one BRAM. 

Because $R_F=2$ might be difficult to achieve for some designs, we can also support fractional frequency ratios through the more complex use-case in Figure \ref{fig:fractional_stream}, where DWC denotes a data width converter. Here we want to co-locate an odd number of buffers and read them back through the 2 ports in a balanced way such that we can clock the memory at $R_F=N_b/2$ times the compute frequency. To achieve this, we split, for example, buffer 1 into two smaller buffers 1 (ODD) and 1 (EVEN) containing the odd and even addresses of the original buffer respectively, resulting in $N_b+1$ total buffers. Crucially, buffers 1 (ODD) and 1 (EVEN) must be allocated to different BRAM ports for readback. 
In operation and without any backpressure applied to any stream, each buffer is read $2/(N_b+1)$ times in each memory cycle except buffer~1, which is being read through both ports and thus gets twice the throughput at $4/(N_b+1)$ reads per memory cycle and $2N_b/(N_b+1)$ reads per compute cycle when $R_F=N_b/2$. Since this value is greater than 1, the compute logic will apply backpressure to stream~1. If the memory streamer has adaptive read slot allocation, it will redistribute the cycles not utilized by stream~1 to other streams, enabling them to meet their throughput requirements as well. Figure \ref{fig:fractional_stream} exemplifies this approach for $N_b=3$. This approach is more complex because it requires additional logic for data width conversion.

\section{Evaluation}\label{sec:Evaluation}

We evaluate our OCM packing approach on two classes of inference accelerator: embedded-class CNN inference accelerators targeting Xilinx Zynq devices, denoted CNV, as well as Resnet-50, denoted RN50. CNV-W1A1 and CNV-W2A2 belong to the BNN-Pynq suite of binarized inference accelerators and target image classification, achieving an accuracy of 79.54\% and 84.8\% respectively on the CIFAR-10 benchmark. 

To support the implementation of our methodology, we modified the FINN HLS Library by adding C++ functions describing the MVAU with streaming weights. To enable the more complex use-cases described in Section \ref{sec:Packing}, we also developed an adaptive weight streamer in Verilog RTL, packaged as a Vivado IP and available on GitHub\footnote{https://github.com/Xilinx/finn/tree/dev/finn-rtllib/memstream}.

We apply the methodology described in \cite{kroes2020pack} to each baseline accelerator to obtain a buffer packing configuration that minimizes OCM requirements. We utilize the same packing algorithm hyperparameters from \cite{kroes2020pack} listed in Table \ref{tab:hyperparams} for each accelerator respectively, where $H_b$ denotes the maximum allowed number of weight buffers packed into each BRAM, $N_p$ denotes the population size of the genetic algorithm, $N_t$ denotes the tournament selection group size, and the remaining parameters are probabilities affecting genetic mutation. For each accelerator, we develop an inter-layer packing solution, where buffers from different resblock convolutional layers are allowed to be packed together into a single BRAM, but in the case of Alveo, only for layers located on the same SLR. Therefore, the packing is floorplan-specific on Alveo devices. We exclude the top and bottom layers from the packing, since the first layer weights are small in size, while the last fully connected layer weight memory is stored in URAM, HBM or DDR, depending on the availability of each resource on the target FPGA platform. The maximum bin height is set to 3 or 4 in our experiments, requiring a memory frequency equal to 1.5x or 2x the compute frequency respectively to maintain the inference throughput.

\begin{table}[]
\small
\caption{\label{tab:hyperparams}Packing GA Hyperparameters}
\centering
\begin{tabular}{ccccccc}
\toprule
\textbf{Accelerator} & $H_B$ & $N_p$ & $N_{t}$ & $P_{adm}^{w}$ & $P_{adm}^{h}$ & $P_{mut}$ \\\midrule
CNV        & 3/4     & 50    & 5       &  0            & 0.1           & 0.3       \\
RN50    & 3/4    & 75    & 5       &  0            & 0.1           & 0.4       \\\bottomrule
\end{tabular}
\end{table}

Table \ref{tab:mem_subsystems} presents the resource utilization characteristics of original and packed memory subsystems of each accelerator. Packed memory subsystems are denoted by the suffix P3 and P4 depending on the maximum bin height allowed in each experiment. For ResNet accelerators which target multiple Alveo cards, an additional suffix indicates the target Alveo device. We observe approximately 20\% increase in OCM utilization efficiency for the CNV-W1A1 from FCMP, while the initial efficiency of CNV-W2A2 is higher and therefore the increase is less. Each of these pays a moderate cost in logic overhead of a few thousand LUTs. For ResNet accelerators, the efficiency gain is more dramatic, from close to 50\% to over 90\% but the LUT overhead is also greater in absolute terms, due to the much larger number of FIFOs required for clock domain crossing, as well as streamer addressing logic. In all cases we observe that using the smaller bin height of 3 yields less OCM efficiency (approximately 5-10\% less than for bin height of 4) but also more logic overhead, due to the more complex weight streamer structure as illustrated in Figure \ref{fig:streamer_sym}. However, the required memory frequency is 25\% lower for bin height 3 compared to bin height of 4, and therefore this style of memory packing may be more suitable for designs which have a relatively high compute frequency.  

\begin{table}[t]
\caption{Packed Memory Subsystems\label{tab:mem_subsystems}}
\centering
\begin{tabular}{cccc}
\toprule
\textbf{\makecell{Accelerator}} & \textbf{\makecell{Logic \\ (kLUT)}} & \textbf{\makecell{BRAMs}} & \textbf{\makecell{E \\ (\%)}} \\ \midrule
\makecell{CNV-W1A1}        &               & 126           &  67.6     \\
\makecell{CNV-W1A1-P3}        & 4.8            & 108           &  78.8     \\
\makecell{CNV-W1A1-P4}        & 3.9            & 96           &  88.7     \\\midrule
\makecell{CNV-W2A2}        &               &  208         &  79.9        \\
\makecell{CNV-W2A2-P3}        & 6.7           &  194         &  85.6        \\
\makecell{CNV-W2A2-P4}        & 1.8           &  188         &  88.4        \\\midrule
\makecell{RN50-W1A2-U250} &                & 2320           &  52.9         \\
\makecell{RN50-W1A2-U250-P3} & 64.9          & 1804           &  68.0         \\
\makecell{RN50-W1A2-U250-P4} & 51.9          & 1632           &  75.3         \\
\makecell{RN50-W1A2-U280-P4} & 38.8          & 1327          &  92.6         \\
\makecell{RN50-W2A2-U250-P4} & 66.5          & 2642           & 92.6       \\\bottomrule
\end{tabular}
\end{table}

We next implement memory-packed as well as folded accelerators in their target platforms to evaluate and compare any loss of throughput. We opt to utilize packing with bin height of 4 for the highest density and lowest LUT overhead. Table \ref{tab:hw_packed} presents the results of implementation for each accelerator and target FPGA combination, where $\delta_{FPS}$ indicates relative throughput reduction from baseline and is defined as $min(F_c,F_m/2)$ divided by the original compute frequency of the baseline, non-packed accelerator. For CNV, we see no problem meeting the frequency requirement for the memory subsystem on the Zynq 7020. Furthermore, we were able to successfully port the CNV-W1A1-P4 accelerator to a smaller Zynq device, the 7012S, without any loss of throughput.

\begin{table}[]
\caption{\label{tab:hw_packed}Comparison of Packed and Folded Accelerators}
\centering
\begin{tabular}{cccccc}
\toprule
\multirow{2}{*}{Accelerator} & \multicolumn{2}{c}{Utilization (\%)} & \multicolumn{2}{c}{Frequency (MHz)} & \multirow{2}{*}{\makecell{$\delta_{FPS}$ \\ (\%)}}\\
                             & LUT              & BRAM              & $F_c$            & $F_m$            &          \\\midrule
CNV-W1A1-7020-P4               & 58               & 50                & 100              & 200              & 0   \\
CNV-W1A1-7012S-P4              & 90               & 97                & 100              & 200              & 0    \\\midrule
RN50-W1A2-U250-P4            & 63               & 62                & 183              & 363            & 12       \\
RN50-W1A2-U280-P4            & 99               & 59                & 138              & 373              & 32     \\
RN50-W2A2-U250-P4            & 94               & 75                & N/A              & N/A              & N/A              \\\midrule
RN50-W1A2-U280-F2            & 61               & 64                & 191              & -              & 51         \\
\bottomrule
\end{tabular}
\end{table}

For RN50 accelerators, timing closure with packed memory subsystems proved to be challenging at the target 200 MHz compute frequency and 400 MHz memory frequency. RN50-W1A2-U250-P4 achieved $R_F=2$ but both clocks failed their targets by approximately 12\%, with a corresponding decrease in inference throughput. RN50-W1A2-U280-P4 almost achieved timing closure for the memory clocks but the compute clock frequency decreased by 32\%. This is likely caused by the very dense design, utilizing 99\% of all LUTs on the U280. As an alternative method of OCM efficiency increase, we also include a folded version of the binary ResNet50 accelerator targeting the Alveo U280 with half the parallelism compared to the baseline, denoted RN50-W1A2-U280-F2. This acclerator achieves similar compute frequency on U280 to the baseline U250 accelerator, but has half the per-cycle throughput and is therefore 51\% slower than the baseline. We can therefore determine that the FCMP approach is 38\% faster than the folding approach for this design and platform combination. Finally, RN50-W2A2-U250-P4 synthesized within the resource limits of the U250, with LUTs becoming the bottleneck, but failed to be placed. This accelerator will require a combination of folding and packing to achieve a working implementation on U250, but crucially, OCM is no longer a bottleneck after memory packing.

\section{Conclusion}

The FCMP methodology presented in this work enables increased memory resource utilization efficiency in modern FPGAs. Our approach is general, i.e. can be applied to any digital circuit utilizing large parameter memories that are read in predictable fashion at run-time. In the specific context of dataflow CNN inference accelerators, where memory resource availability is often a design bottleneck, our technique can enable a specific accelerator design to target more devices by becoming more efficient in its block RAM usage. Furthermore, FCMP allows a more fine-grained resource-throughput trade-off compared to other alternatives such as accelerator folding. However, our experiments have shown that achieving timing closure at the desired frequency multiples for the memory subsystem is not always trivial, although in practice it is easier than initially expected, especially for monolithic FPGA devices. For multi-die FPGAs, the design process is complicated by the requirement for explicit floorplanning, and future work could focus on integrating the memory packing approach into a design space exploration framework to perform automatic floorplanning or partitioning of a design in the context of either multi-SLR or multi-FPGA systems. An alternative avenue for future work is to extend the concepts presented here to increase the OCM utilization efficiency of other parts of dataflow CNN accelerators, such as activation storage.



\bibliographystyle{IEEEtran}

\bibliography{references}

\end{document}